\documentclass[letterpaper, 10 pt, conference]{ieeeconf} 
\usepackage{amsmath}
 \usepackage{graphicx}
  \usepackage{float}
  \usepackage{color}

 \include{mathsym}
\newenvironment{braced}
 {\par\smallskip\hbox to\columnwidth\bgroup
  \hss$\left\{\begin{minipage}{\columnwidth}}
 {\end{minipage}\right.$\hss\egroup\smallskip}

\usepackage{enumerate}  
 \graphicspath{ {images/} }
 
\IEEEoverridecommandlockouts                              

\title{\LARGE \bf
Gait Generation using Intrinsically Stable MPC\\ in the Presence of Persistent Disturbances
}

\author{Filippo M. Smaldone, Nicola Scianca, Valerio Modugno, Leonardo Lanari, Giuseppe Oriolo
\thanks{$^{1}$The authors are with the Dipartimento di Ingegneria Informatica, Automatica e Gestionale, Sapienza Universit\`a di Roma, via Ariosto 25, 00185 Roma, Italy. E-mail: \{scianca,modugno,lanari,oriolo\}@diag.uniroma1.it}%
}

\begin{document}

\maketitle
\thispagestyle{empty}
\pagestyle{empty}

\begin{abstract}
Maintaining balance while walking is not a simple task for a humanoid robot because of its complex dynamics. The presence of a persistent disturbance makes this task even more challenging, as it can cause a loss of balance and ultimately lead the the robot to a fall. In this paper, we extend our previously proposed Intrinsically Stable MPC (IS-MPC), which guarantees boundedness of the CoM with respect to the ZMP, to the case of persistent disturbances. This is achieved by designing a disturbance observer whose 
estimate is used to compute a modified stability constraint included in the QP problem formulation. The method is validated by MATLAB simulations for the LIP as well as dynamic simulations for a NAO humanoid in DART.
\end{abstract}

\section{Introduction}

Interest in humanoid robotics has considerably increased in the last decade, leading to major improvements both on the constructive side, with more reliable platforms being available, and on the control side.

Maintaining balance while walking is still a challenging task. The problem is usually approached by controlling the position of the Zero Moment Point (ZMP)~\cite{KaHiHaYo:14}, i.e., the  point on the ground for which the horizontal components of the contact moments become zero. In order to guarantee dynamic balance, the ZMP has to be kept inside the support polygon of the robot at all times during the gait. Most position controlled humanoids adapt the ZMP through the Center of Mass (CoM) of the robot.

The dynamics relating the ZMP to the CoM is nonlinear, however standard simplifying  assumptions lead to the well-known Linear Inverted Pendulum (LIP) model which is the basis for many popular approaches to humanoid gait generation. The influential paper~\cite{KaKaKaFuHaYoHi:03} uses the Cart Table (CT), inverse of the LIP, and derives a preview controller. Constraints were added in~\cite{Wi:06} using a Model Predictive Control (MPC) scheme. This formulation was also extended in order to perform automatic footstep placement in~\cite{HeDiWiDiMoDi:10}.

For humanoids, since the area of the footprint is small, a push or an external persistent force may easily throw the robot off balance. The first case, push recovery, has been addressed for example with the Capture Point~\cite{PrCaDrGo:06} and MPC schemes have shown to be also very effective~\cite{Wi:06,ScCoDeLaOr:16} just to cite a few. However disturbances and uncertainties can have a detrimental effect on an MPC-controlled system, if not properly accounted for. One possibility is to adopt a robust approach and design a controller that is able to withstand bounded unknown disturbances~\cite{ChRoZa:01, MaSeRa:05, PaRaWr:11}. In the context of humanoid locomotion, these ideas have been developed in~\cite{ViWi:17} leading to the precise computation of safety margins to cope with a given set of uncertainties.

When dealing with strong persistent disturbances, this approach may result very conservative. Another possibility is to introduce a disturbance observer and design a controller which is able to counteract the disturbance. An first notable example is~\cite{KaKaMoYoLa:12} where the external force is estimated evaluating its effect on the humanoid. Other examples include \cite{CzKeUr:09} which incorporates an observer in a preview formulation, or approaches based on the Divergent Component of Motion such as~\cite{EnMeOt:17} and~\cite{GrLeAs:16}. Similar ideas are employed to deal with unmodeled dynamics during walking, such as in~\cite{HaEr:13} and~\cite{SaOh:08}.


\begin{figure}
\centering\includegraphics[height=0.7\columnwidth]{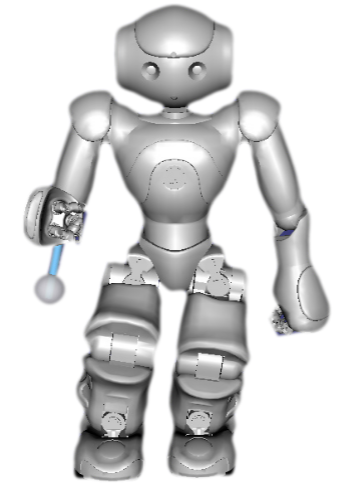}
\caption{A pendulum attached to the robot elbow is used to simulate an unknown persistent disturbance acting on the humanoid while walking.}
\label{fig:NAOPendulum}
\end{figure}

In~\cite{ScCoDeLaOr:16} we proposed an MPC scheme, adding a {\em stability constraint}, in which the CoM trajectory is guaranteed to be bounded with respect to the ZMP. The resulting scheme is defined as Intrinsically Stable MPC (IS-MPC). The control scheme was further developed in~\cite{ScDeLaOr:19} where we also proved how, using available preview information (e.g. coming from a footstep planner), the stability constraint can make the MPC scheme recursively feasible, which means that feasibility (i.e., the existence of a solution under the imposed constraints) at one iteration implies feasibility at the next iteration.

In this paper we extend the stability constraint of the IS-MPC to counteract the effect of a persistent disturbance (e.g., a humanoid carrying an unknown moving payload as illustrated in Fig.~\ref{fig:NAOPendulum}). This is done with the aid of an observer and tested in a range of different situations using dynamic simulations.

The paper is organized as follows.
In the next section we briefly recall IS-MPC in the absence of  disturbances (nominal case). Section~\ref{sec:PerturbedCase} introduces the perturbed LIP model. The ideal case of known disturbance together with the corresponding modified stability constraint are discussed in Sect.~\ref{sec:knownDisturbance}. Removing the known disturbance hypothesis, a disturbance observer is introduced in Sect.~\ref{sec:ObsMPC} together with its use in the resulting observer-based IS-MPC. Simulations on the LIP and dynamic simulations on a NAO humanoid robot are also performed to validate the proposed approach. Section \ref{sec:conclusions} offers a few concluding remarks.

\section{IS-MPC: The Nominal Case}
\label{sec:nominalISMPC}

In this section we provide a brief review of the IS-MPC gait generation method. See~\cite{ScCoDeLaOr:16,ScDeLaOr:19} for further details.

Assume that the humanoid is walking on flat horizontal ground, and denote the position of the humanoid Center of Mass (CoM) and Zero Moment Point (ZMP) as $(x_c,y_c,z_c)$ and $(x_z,y_z,0)$, respectively.
The dynamic equation relating the CoM and the ZMP can be derived by balancing moments around the ZMP, e.g., see~\cite{WiTeKu:16}. Assuming the CoM height $z_c$ to be constant and neglecting angular momentum contributions around the CoM leads to the Linear Inverted Pendulum (LIP) model, where the $x$-axis (sagittal) and $y$-axis (coronal) dynamics are linear, identical and decoupled. For illustration, consider only the sagittal motion
\begin{equation}
\ddot{x}_c=\eta^2(x_c-x_z),
\end{equation}
with $\eta=\sqrt{g/\bar z_c}$, where $g$ is the gravity acceleration and $\bar z_c$ is the constant CoM height. Note that the ZMP position $x_z$ acts as an input in this model.

The LIP is decomposed into a stable and an unstable subsystem by using the following change of coordinates:  
\begin{equation}
\begin{split}
{x}_u&=x_c + \dot x_c /\eta\\
{x}_s&=x_c - \dot x_c /\eta. 
\end{split}
\label{eq:dec}
\end{equation}
The dynamics of $x_u$, also known as {\em divergent component of motion}~\cite{TaMaYo:09} or {\em capture point}~\cite{PrCaDrGo:06}, is
\begin{equation}\label{eq:XuDynamics}
\dot x_u = \eta\, (x_u - x_z).   
\end{equation}
Although this dynamics is unstable, $x_u$ (and hence $x_c$) will not diverge with respect to $x_z$ provided that 
\begin{equation}
x_u^k =  \eta \int_{t_k}^\infty e^{-\eta(\tau -t_k)}x_z(\tau)d\tau.
\label{eq:boundedness}
\end{equation}
Equation~(\ref{eq:boundedness}), called the {\em stability condition} in the following, is a relationship between 
the value of $x_u$ at the current time $t_k$, denoted by $x_u^k$, and the {\em future} values of the input $x_z$, and is therefore non-causal~\cite{LaHuMa:14}. 

Intrinsically Stable MPC (IS-MPC) is a scheme for humanoid gait generation that uses a causal stability constraint derived from condition~(\ref{eq:boundedness}) in order to guarantee that the gait is internally stable, i.e., that the CoM remains bounded with respect to the ZMP. The prediction model is a dynamically extended LIP
\begin{equation}
\left( \begin{array}{c}
\dot x_c\\
\ddot x_c\\
\dot x_z
  \end{array} \right) = 
\left( \begin{array}{ccc}
0& 1&  0\\
\eta^2& 0&  -\eta^2\\
0& 0 & 0
  \end{array} \right)
\left( \begin{array}{c}
x_c\\
\dot x_c\\
x_z
\end{array} \right) + \left( \begin{array}{c} 0 \\ 0 \\ 1 \end{array}\right) \dot x_z,
\label{eq:ThirdOrder}
\end{equation}
with the ZMP velocity $\dot x_z$ now acting as input. IS-MPC uses piecewise-constant inputs, i.e., $\dot x_z(t) = \dot x_z^i$ for $t\in [t_i, t_{i+1}]$, with $t_{i+1}-t_i=\delta$ the duration of sampling intervals.
The MPC {\em control horizon} is $C\cdot\delta$.

Although IS-MPC can perform automatic footstep placement (AFS), in this paper we will for simplicity consider the case of given footsteps (a simulation with AFS is included however in Section~\ref{sec:SimulationsLip}). 
In this case, only the ZMP  and the stability constraints must be enforced.

The ZMP constraint guaranteeing dynamic balance is expressed as
\begin{equation}
R_j^T \left( \begin{array}{c} \delta\sum_{l=k}^{k+i-1}\dot x_z^{l} - x_f^j \\ [10pt]
\delta\sum_{l=k}^{k+i-1}\dot y_z^{l} - y_f^j \end{array}  \right) \leq
\frac{1}{2}\left(  \begin{array}{c} f_x \\  [5pt]
f_y \end{array}  \right) -
R_j^T \left(  \begin{array}{c} x_z^{k} \\ [5pt]
y_z^{k} \end{array}  \right),
\label{eq:ZMPcon}
\end{equation}
where $R_j^T$ is the rotation matrix associated to the orientation of the $j$-th footstep, $(x_f^j,y_f^j)$ is its position, and $f_z$, $f_y$ are the dimensions of a rectangular region approximating the footprint. The above is the expression of the constraint during single support; the double support constraint can be expressed in a similar way.

The {\em stability constraint} is derived from~(\ref{eq:boundedness}) using the fact that the ZMP is piecewise-linear:
\begin{equation}
\sum_{i=0}^{C-1} e^{-i\eta\delta}\dot x_z^{k+i} = -\sum_{i=C}^\infty e^{-i\eta\delta}\dot{\tilde x}_z^{k+i} + 
\frac{\eta}{1-e^{-\eta\delta}}(x_u^k - x_z^k).
\label{eq:StabConstr}
\end{equation}
Here, the left-hand side gathers the ZMP velocities $\dot x_z^{k},\dots,\dot x_z^{k+C-1}$ within the control horizon, which are are the MPC decision variables. The right-hand side depends on the system state at $t_k$ as well as on the {\em tail}, i.e., the conjectured values $\dot{\tilde x}_z^{k+C}, \dot{\tilde x}_z^{k+C+1}, \dots$ of the ZMP velocities {\em after} the control horizon. This conjecture, which is needed to obtain a causal constraint, can be made using the available preview information on the footstep plan ({\em anticipative tail}). More details on ZMP velocity tails are given in~\cite{ScDeLaOr:19}.

Collecting the MPC decision variables in 
\begin{eqnarray*}
\dot X_z^k &=& (\dot x_z^k \>\> \ldots \>\> \dot x_z^{k+C-1})^T\\
\dot Y_z^k &=& (\dot y_z^k \>\> \ldots \>\> \dot y_z^{k+C-1})^T,
\end{eqnarray*}
the generic MPC iteration at $t_k$ consists in solving the following Quadratic Programming (QP) problem:

\medskip
\begin{braced}
\[
\min_{\dot X_z^k,\dot Y_z^k}  
\|\dot X_z^{k} \|^2  + \|\dot Y_z^{k} \|^2 
\]
\centerline{subject to:}

\begin{itemize}
\item ZMP constraints (\ref{eq:ZMPcon});
\item stability constraints (\ref{eq:StabConstr}) for $x$ and $y$.
\end{itemize}
\end{braced}
\medskip

Once the problem is solved, the first sample $\dot x_z^k$ of the optimal input sequence is used to integrate the LIP dynamics along $x$ (analogously for $y$). This results in a CoM reference trajectory that can be tracked by the humanoid robot using a standard kinematic controller.

\section{The Perturbed Model}
\label{sec:PerturbedCase}

Consider now a disturbance\footnote{In general, the disturbance will be a vector $(d_x,d_y)$ and will include a component acting along $y$. However, in the following we focus on the $x$ dynamics and therefore we shall simply write $d$ in place of $d_x$.} $d$ acting on the LIP dynamics as follows:
\begin{equation}
\ddot{x}_c=\eta^2(x_c-x_z)+d.
\label{model_plus_dist}
\end{equation}
This disturbance can represent external forces acting as well as unmodeled dynamics and uncertainties. For example, starting from the general CoM dynamics~\cite{ViWi:17} one can write $d$ as
\begin{equation}
d_x = \left(1 - \eta^2 \frac{z_c}{\ddot z_c + g}\right) \ddot x_c - \eta^2 \frac{\dot L_y}{m(\ddot z_c + g)} + \frac{F_{\rm ext}}{m},
\label{eq:mism}
\end{equation}
where $m$ is the total mass of the robot, $L_y$ its centroidal angular momentum around the $y$-axis and $F_{\rm ext}$ is (the $x$ component of) the resultant of external forces acting on the CoM. The first two terms of this expression are unmodeled dynamics, whereas the third is an actual disturbance.

The prediction model~(\ref{eq:ThirdOrder}) is therefore modified to include the disturbance as
\begin{equation}
\left(\! \begin{array}{c}
\dot x_c\\
\ddot x_c\\
\dot x_z
  \end{array} \!\right) \!=\! 
\left(\! \begin{array}{ccc}
0& 1&  0\\
\eta^2& 0&  -\eta^2\\
0& 0 & 0
  \end{array}\! \right) \!\!
\left( \!\begin{array}{c}
x_c\\
\dot x_c\\
x_z
\end{array} \!\right) 
+ \left( \!\begin{array}{c} 0 \\ 0 \\ 1 \end{array}\!\right) \dot x_z
+ \left( \!\begin{array}{c} 0 \\ 1 \\ 0 \end{array}\!\right) d.
\label{eq:ThirdOrderWithD}
\end{equation}
Applying the same change of variables~(\ref{eq:dec}), the unstable component $x_u$ is now found to be affected by $d$: 
\[
\dot x_u = \eta \, (x_u - x_z)+d/\eta.   
\]
To guarantee boundedness of the CoM with respect to the ZMP in the perturbed case, the stability condition~(\ref{eq:boundedness}) must be modified accordingly:
\begin{equation}\label{initialond new}
x_u^k =  \eta \int_{t_k}^\infty e^{-\eta(\tau -t_k)}x_z(\tau)d\tau - \frac{1}{\eta}\int_{t_k}^\infty e^{-\eta(\tau -t_k)}d(\tau)d\tau.
\end{equation}
A causal implementation of this condition would require, in addition to the conjecture on the ZMP velocities after the control horizon, also knowledge of $d$ from $t_k$ up to infinity. In the next section, we will temporarily assume that knowledge is indeed available in order to devise an IS-MPC for the perturbed case. This hypothesis will be removed in Sect.~\ref{sec:ObsMPC} by introducing a disturbance observer to be used for implementing the stability constraint.

\section{IS-MPC: The Known Disturbance Case}
\label{sec:knownDisturbance}


Assume that the disturbance $d$ is known over $[t_k, \infty)$. From~(\ref{initialond new}) we can then derive the following computable expression of the stability constraint (compare with~(\ref{eq:StabConstr}))
\begin{equation}
\sum_{i=0}^{C-1} e^{-i\eta\delta}\dot x_z^{k+i} \! = \! -\sum_{i=C}^\infty e^{-i\eta\delta} \dot{\tilde x}_z^{k+i} +\frac{\eta}{1-e^{-\eta\delta}}(x_u^k - x_z^k + \Delta_d^k),
\label{eq:BBDD_k}
\end{equation}
having denoted by $\Delta_d^k$ the correction term due to the disturbance
\begin{equation}\label{eq:DeltaD}
\Delta_d^k = \frac{1}{\eta}\int_{t_k}^\infty e^{-\eta(\tau -t_k)}d(\tau)d\tau.
\end{equation}
Consistently with the assumption made for $x_z$ in Sect.~\ref{sec:nominalISMPC}, suppose that the disturbance is piecewise-linear 
\begin{equation}
d(t) =  d^i+\dot{d}^i(t-t_i), \qquad t\in[t_i,t_{i+1})  
\end{equation}
with $d^i=d(t_i)$. Then, a simple computation gives 
\begin{equation}
\Delta_d^k =
\frac{1-e^{-\eta\delta}}{\eta^3}\sum_{i=0}^\infty e^{-i\eta\delta} \dot{d}^{k+i}+\frac{d^k}{\eta^2}.
\label{eq:deltaDpiecewise}
\end{equation}
Replacing constraint~(\ref{eq:StabConstr}) in the QP formulation with constraint~(\ref{eq:BBDD_k}), where $\Delta_d^k$ is given by~(\ref{eq:deltaDpiecewise}), and similarly for $y$, leads to an IS-MPC scheme where the control inputs (the ZMP velocities within the control horizon) are directly modified by the profile of the disturbance, realizing a form of indirect {\em disturbance compensation}\footnote{Direct compensation is not possible for system~(\ref{eq:ThirdOrderWithD}) because the control input and the disturbance act at different levels.}. In particular, recursive feasibility will be achieved if sufficient preview information is available, and in turn this will guarantee internal stability~\cite[Props.~5 and 6]{ScDeLaOr:19}.

\begin{figure}
\centering
\includegraphics[height=0.7\columnwidth]{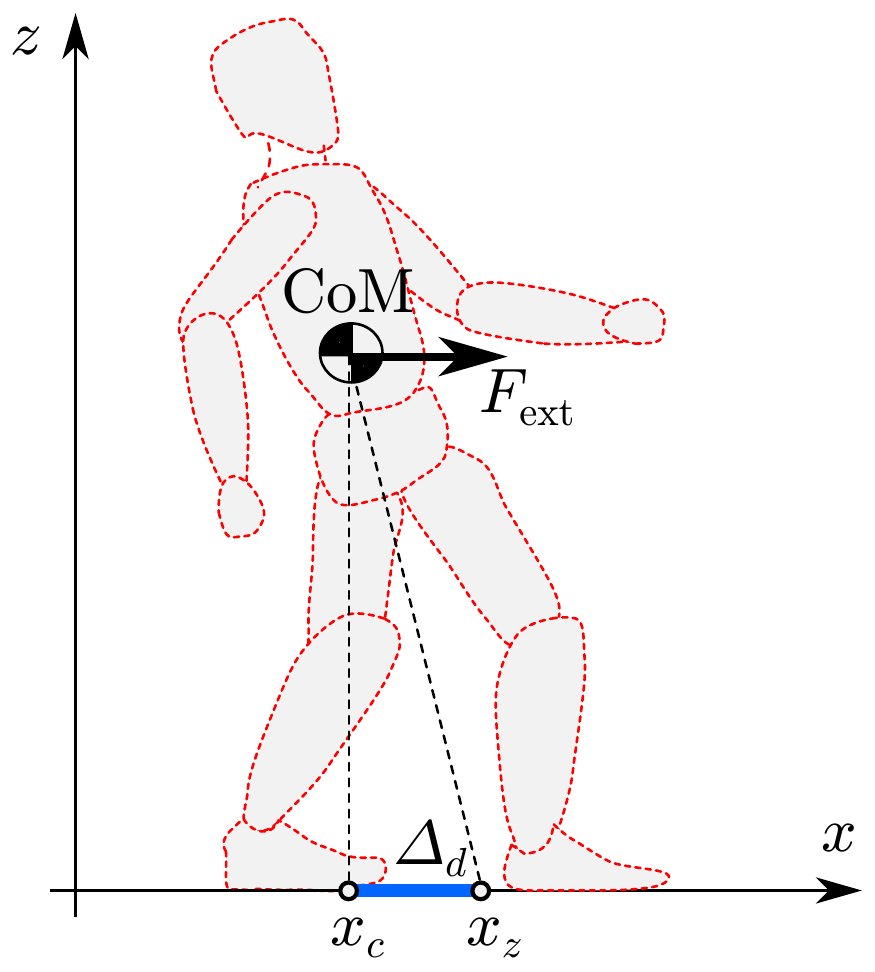}
\caption{Balancing in the presence of a known constant force acting on the CoM: IS-MPC with disturbance compensation produces a steady-state displacement between the CoM and the ZMP that can be interpreted as the humanoid "leaning against" the force. This displacement is exactly equal to the disturbance-related term in the stability constraint.}
\label{fig:LIP_ss}
\end{figure}

To appreciate the effect of the compensation, consider the special case in which the humanoid must balance (i.e., footsteps are fixed) in the presence of a constant disturbance $\bar d = {\bar F}_{\rm ext}/m$, arising from a constant force ${\bar F}_{\rm ext}$ pushing on the CoM. Under the action of IS-MPC with disturbance compensation, the robot converges to a steady state where --- consistently with eq.~(\ref{model_plus_dist}) --- the displacement between the CoM and the ZMP is
\[
x_c - x_z = \frac{{\bar F}_{\rm ext}}{m\,\eta^2}. 
\]
This can be interpreted as the humanoid ``leaning against'' the force in order to counteract it (see Fig.\ref{fig:LIP_ss}). Interestingly,  eq.~(\ref{eq:deltaDpiecewise}) in this case readily provides
\[
\Delta_d^k = \Delta_d = \frac{\bar d_k}{\eta^2} = \frac{{\bar F}_{\rm ext}}{m\,\eta^2},
\]
showing that the correction term due to the disturbance in the stability constraint~(\ref{eq:BBDD_k}) is exactly equal to the steady-state CoM-ZMP displacement.


\begin{figure}
\centering
\includegraphics[width=1\columnwidth]{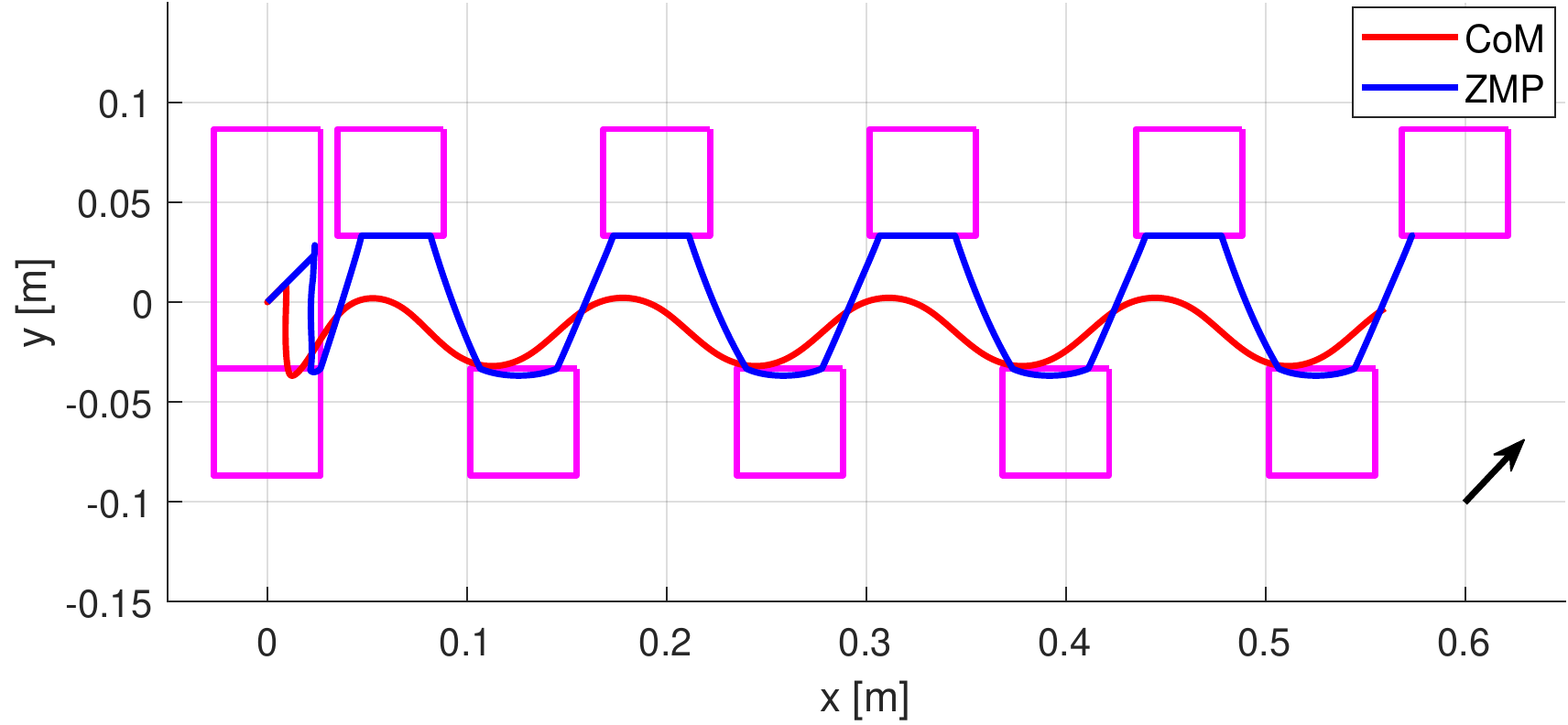}
\includegraphics[width=1\columnwidth]{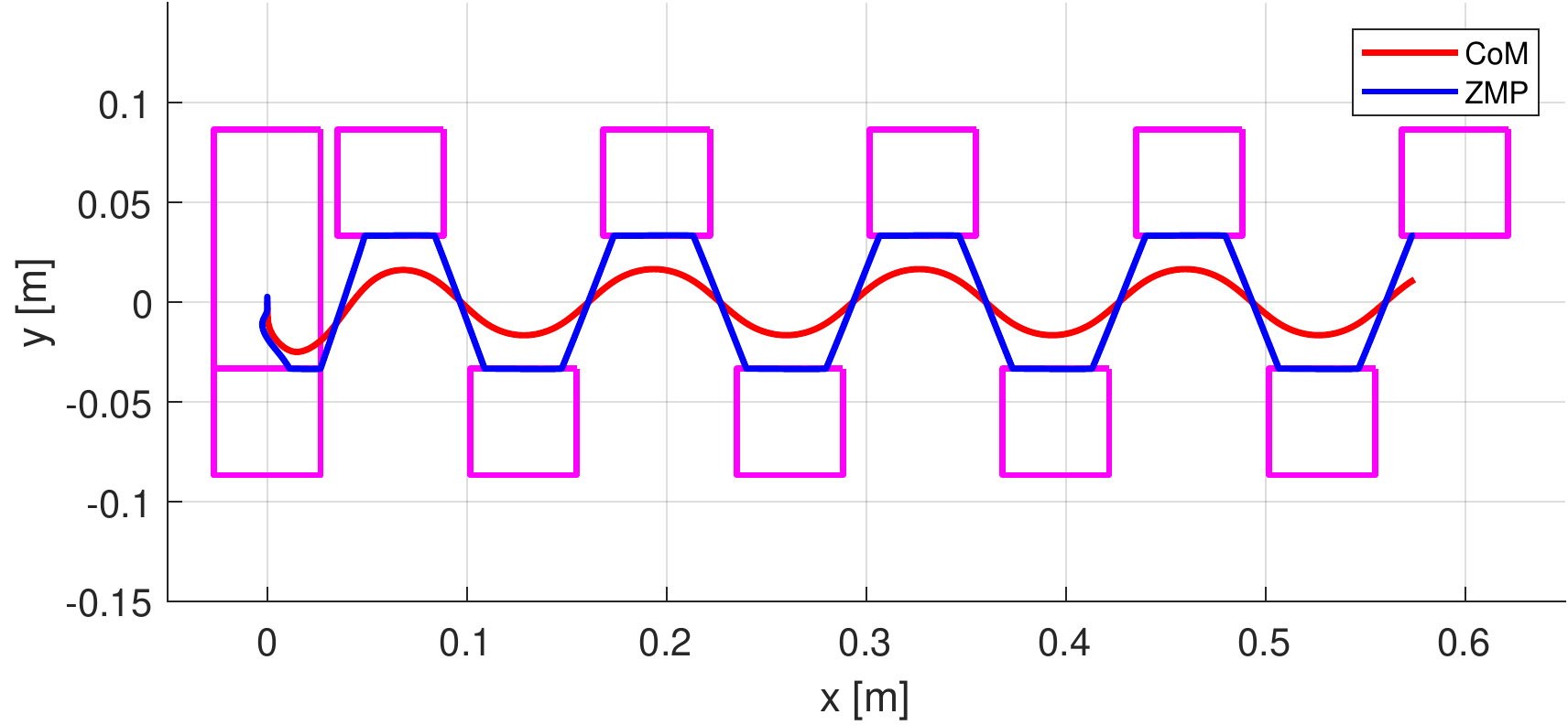}
\caption{Gait generation in the presence of a known constant force acting on the CoM: result of IS-MPC with disturbance compensation (top). Note the arrow indicating the direction of the force. For comparison, the gait produced by IS-MPC when no disturbance acts on the system is also shown (bottom).}
\label{fig:Comparison}
\end{figure}

A similar compensation effect occurs when walking. Figure~\ref{fig:Comparison} shows a gait generated using IS-MPC with disturbance compensation in the presence of a constant force acting on the CoM, in comparison with the gait generated by IS-MPC in the absence of disturbance. The simulation is run in MATLAB with {\tt quadprog} as QP solver, and
uses the following parameters: $m=4.5$~kg, $\bar z_c = 0.33$~m, $f_x=f_y=0.05$~m, duration of the single and double support phases 0.2 and 0.3~s, respectively, $\delta=0.01$~s, $C = 100$; in the perturbed case, the external force along the $x$ axis is 1.8~N, corresponding to a CoM acceleration of 0.4~m/s$^2$, and the same along the $y$ axis. Again, observe how the robot is leaning against the disturbance, as the CoM trajectory is pushed in the opposite direction to the force.

Overall, the behavior resulting from IS-MPC with disturbance compensation can be interpreted as a rather natural {\em anticipative} action aimed at counteracting the effect of the disturbance.

\section{Observer-based IS-MPC}
\label{sec:ObsMPC}

The previous assumption of complete knowledge of the disturbance can be justified in some special cases (e.g., when walking on an inclined plane of known slope), but will not be verified in general. To this end, we design in this section a disturbance observer and discuss its use within an IS-MPC scheme with disturbance compensation. We will then showcase the performance of the resulting gait generation method via simulations on the LIP model and dynamic simulations on the NAO humanoid robot.

\subsection{Disturbance observer}
\label{sec:disturbanceObserver}


In general, the value of the disturbance $d$ is unknown. However, we can estimate it from other measurements; in particular, in the following we assume that the coordinates of the CoM and the ZMP, respectively $x_c$ and $x_z$, are measured (this is a rather standard occurrence in humanoids). Since $d$ is piecewise-linear (see Sect.~\ref{sec:knownDisturbance}), we can adopt the following disturbance model (exosystem):
\[
\ddot d = 0,
\]
and use it to extend the perturbed model~(\ref{eq:ThirdOrderWithD}), obtaining a system with state $x=(
x_c,\dot{x}_c,x_z,d,\dot{d})$ and characterized by the following matrices
\begin{equation}
\begin{split}
A=&
\left( \begin{array}{ccccc}
0& 1&  0& 0 &0\\
\eta^2& 0&  -\eta^2& 1 & 0\\
0&0&0&0&0\\
0& 0& 0& 0& 1\\
0& 0& 0& 0& 0\\
  \end{array} \right)\ \ \ B=
\left( \begin{array}{c}
0\\
0\\
1\\
0\\
0\\
  \end{array} \right)\\
  C =& \left( \begin{array}{ccccc}
1& 0& 0& 0 & 0\\
0& 0& 1& 0 & 0\\
  \end{array} \right).
  \end{split}
\label{eq:ThirdOrderWithDExt}
\end{equation}
Since the system is easily found to be observable, we can build an asymptotic observer
\begin{equation}
\dot{\hat{x}}=A \hat x +Bu+G(C\hat x- y)
\label{eq:observer}
\end{equation}
where $\hat{x}$ is the observer state and $y$ are the available measurements. The  gain matrix $G$ can be computed by simple pole placement. This observer is guaranteed to reconstruct asymptotically any piecewise-linear disturbance signal. 

\subsection{Observer-based stability constraint}

To perform IS-MPC with disturbance compensation in the general case when $d$ is unknown, the estimate $\hat d$ provided by the asymptotic observer~(\ref{eq:observer}) can be
used in the stability constraint~(\ref{eq:BBDD_k}--\ref{eq:DeltaD}). Since the observer only produces the current value $\hat d^k$ of $\hat d$ and does not perform any kind of prediction, the straightforward choice is to compute the correction term $\Delta_d^k$ by replacing $d(\tau)$ with $\hat d^k$ in the integral:
\begin{equation}
\Delta_d^k = \frac{\hat{d}^k}{\eta^2}.
\label{eq:approx}
\end{equation}
While this is obviously an approximation, it should be considered that in the MPC algorithm $\Delta_d^k$ is recomputed at each sampling instant; as a consequence,  observer-based IS-MPC can provide compensation also for slowly-varying signals. This will be shown via simulations in the remainder of this section, which also discusses some additional ideas for achieving compensation of a larger class of disturbances.

\subsection{Simulations on the LIP}
\label{sec:SimulationsLip}

We describe now some MATLAB simulations of the perturbed LIP under the action of observer-based IS-MPC. The parameters are exactly the same of the simulation in Sect.~\ref{sec:knownDisturbance}.

\begin{figure}[t]
\centering
\includegraphics[width=\columnwidth]{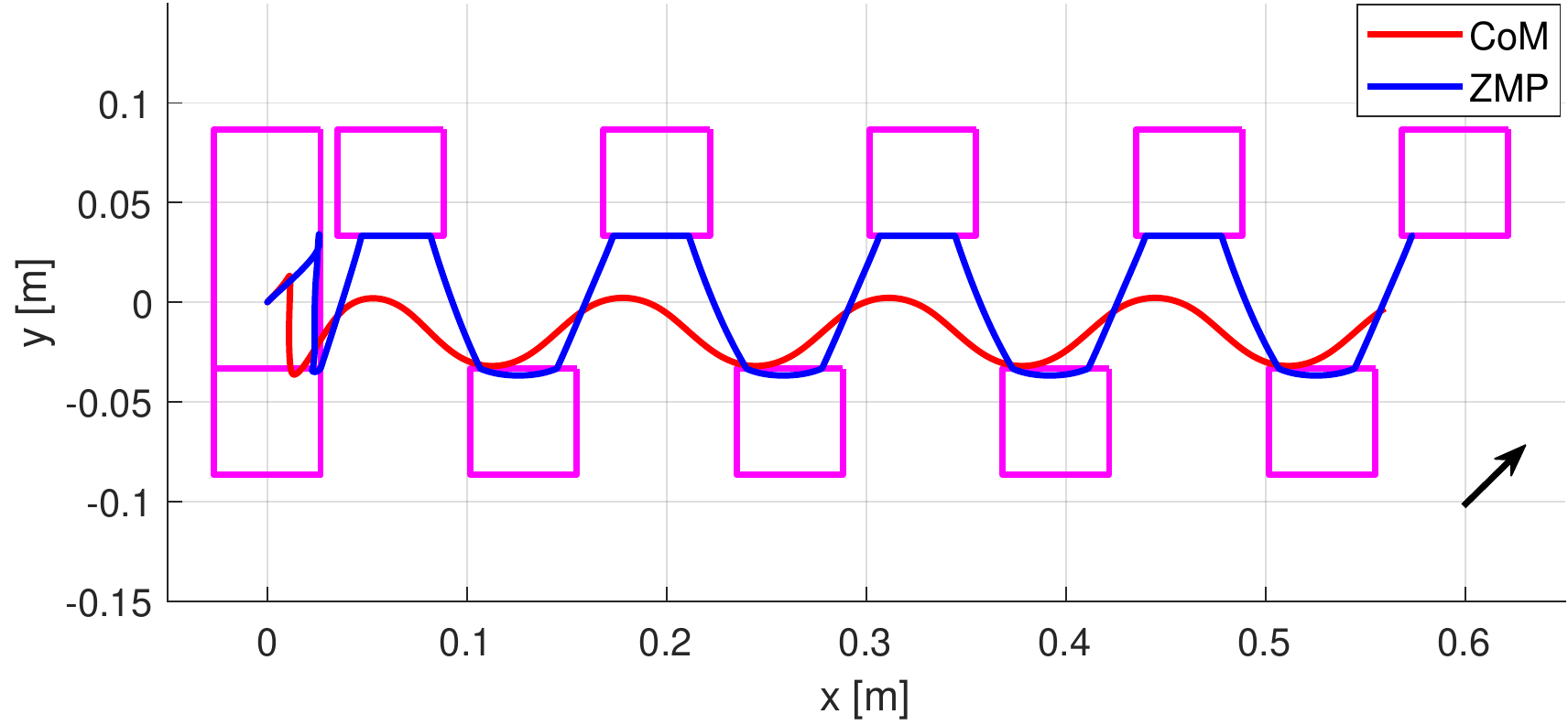}
\caption{Gait generation in the presence of an unknown constant disturbance acting on the CoM: result of observer-based IS-MPC.}
\label{fig:lipSim1}
\end{figure}

In the first simulation, shown in Fig.~\ref{fig:lipSim1}, the LIP is subject to the same constant disturbance of Fig.~\ref{fig:Comparison}, i.e., $\bar d=0.4$~m/s$^2$ on both $x$ and $y$. The observed disturbance $\hat d$ converges therefore to the actual value $\bar d$. As a consequence, the resulting gait is almost indistinguishable from that generated by IS-MPC when the disturbance is known (compare with Fig.~\ref{fig:Comparison}).


\begin{figure}[t]
\centering
\includegraphics[width=\columnwidth]{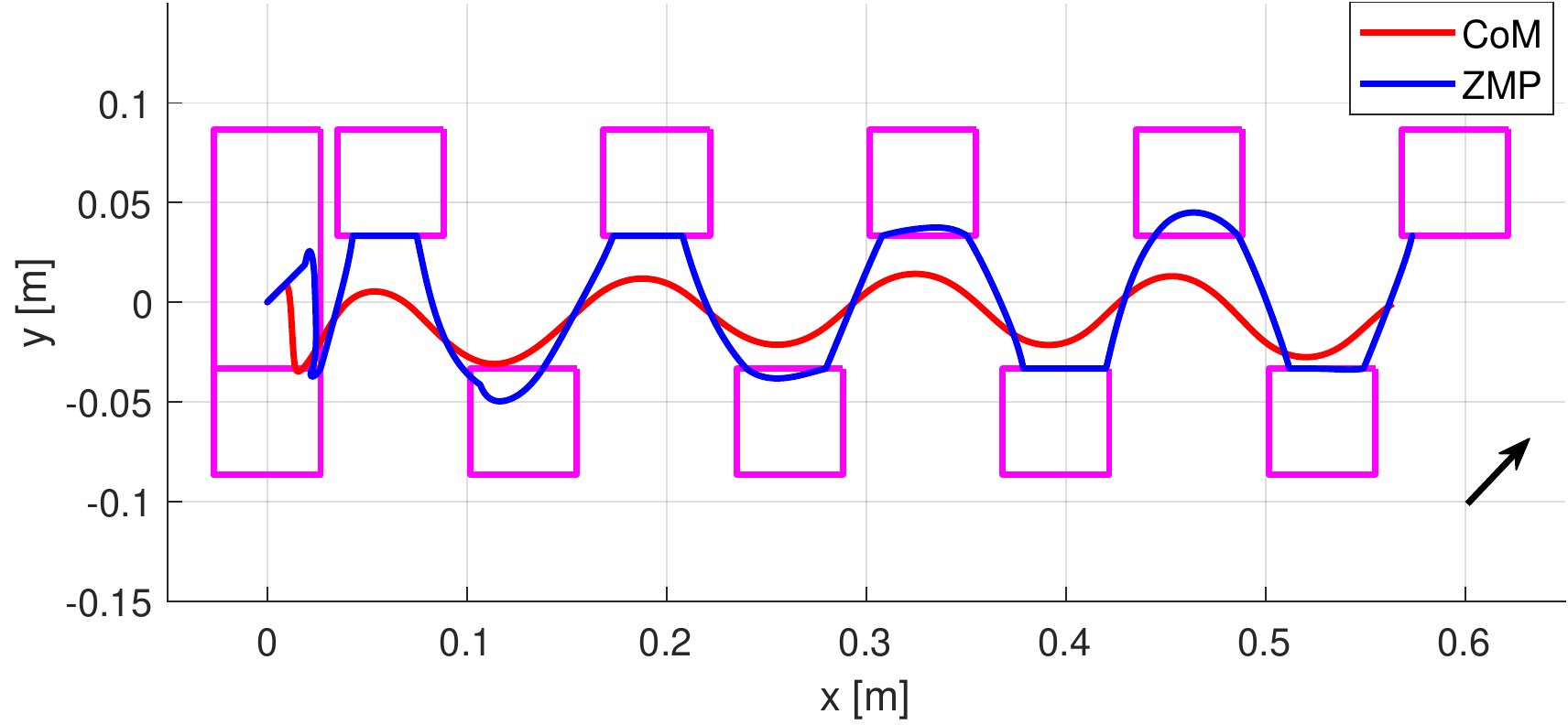}
\caption{Gait generation in the presence of an unknown slowly-varying disturbance acting on the CoM: result of observer-based IS-MPC.}
\label{fig:lipSim2}
\end{figure} 

In the second simulation, we add on both $x$ and $y$ the disturbance signal $d(t) =0.2+0.15\sin (0.45 \pi t)$~m/s$^2$, which is outside the piecewise-linear family due to the sinusoidal term. As shown in Fig.~\ref{fig:lipSim2}, observer-based IS-MPC is still able to produce a stable gait. This proves that the proposed method is robust to two distinct sources of discrepancy: (1) the fact that the observer cannot provide an asymptotically exact estimate of $d$ and (2) the use of the constant value $\hat d^k$ in the stability constraint. 

\begin{figure}[t]
\centering
\includegraphics[width=\columnwidth]{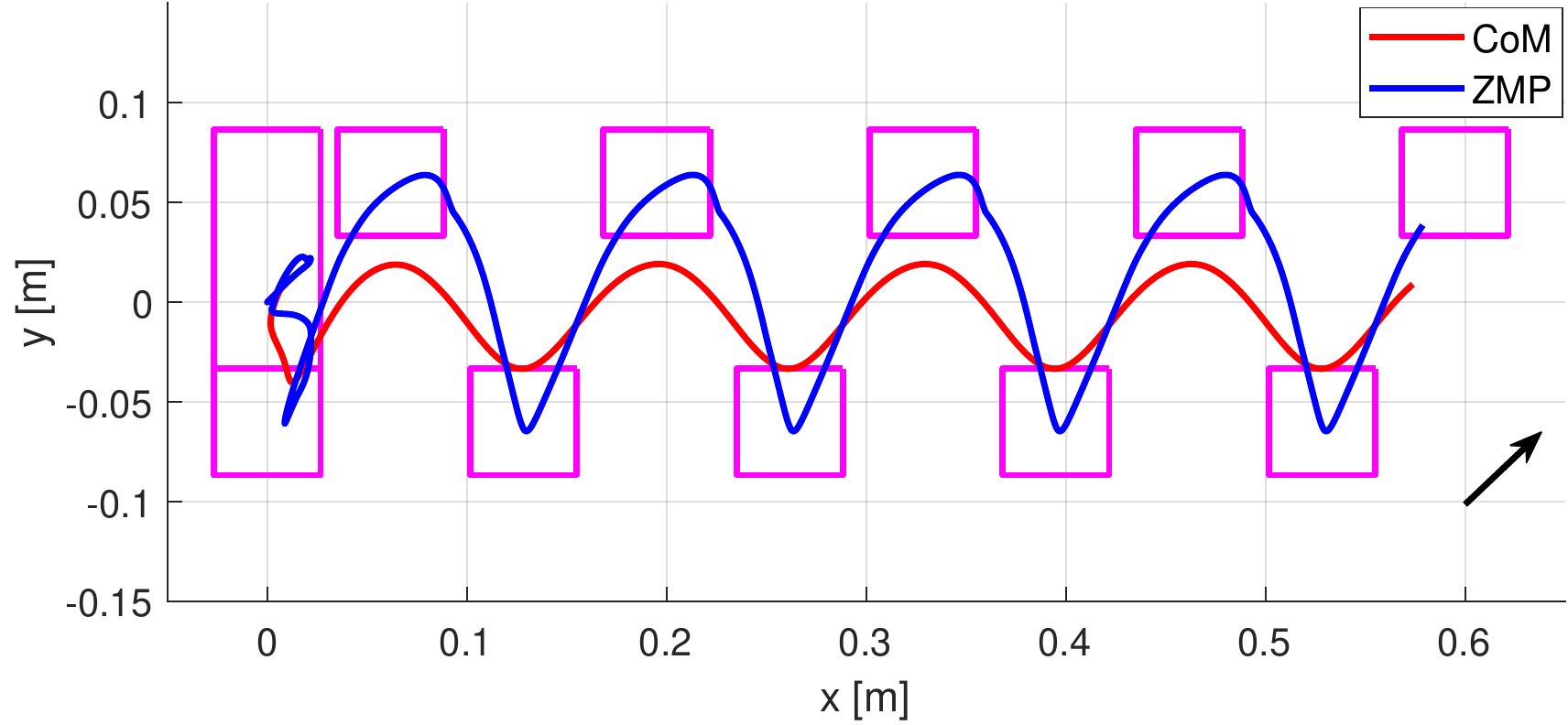}
\includegraphics[width=\columnwidth]{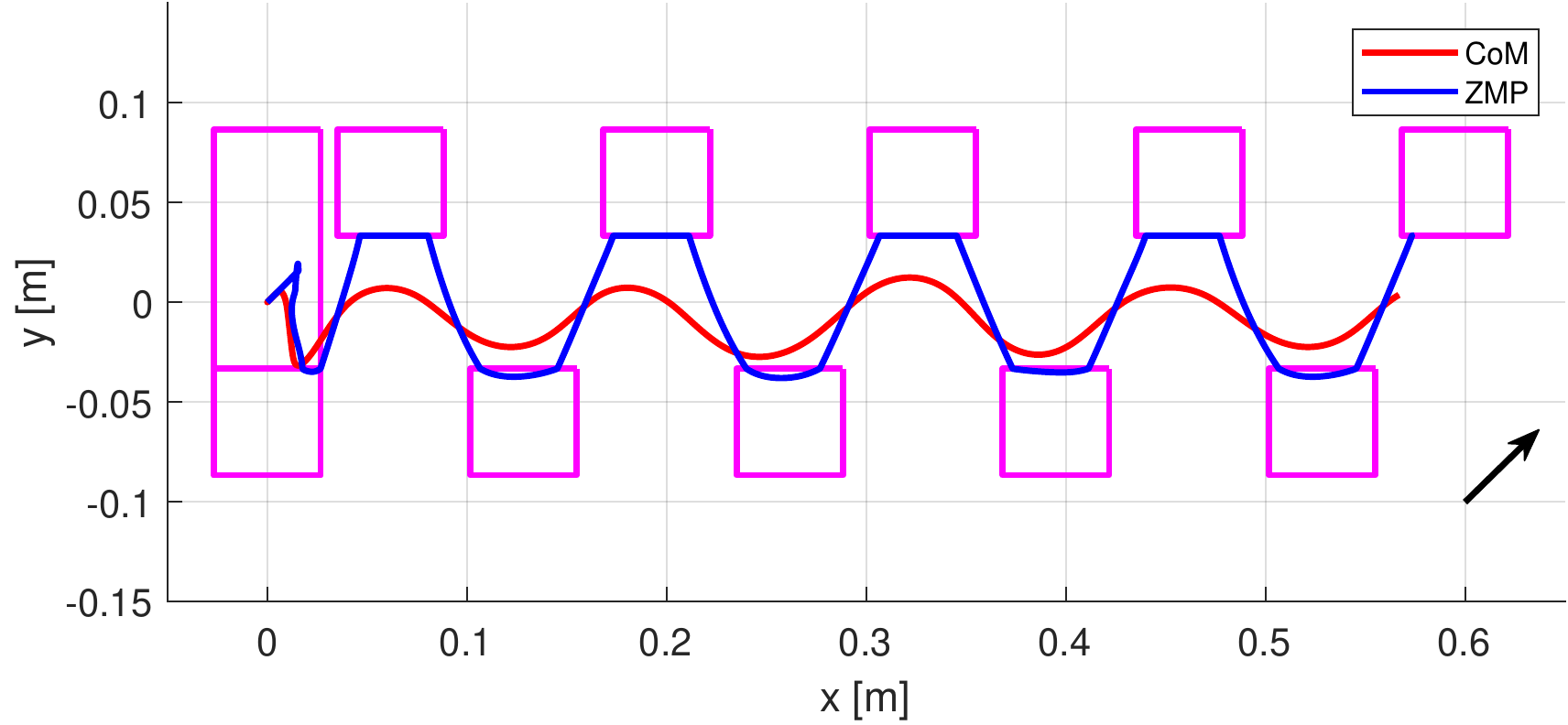}
\caption{Gait generation in the presence of an unknown rapidly-varying disturbance acting on the CoM: result of observer-based IS-MPC with ZMP constraint restriction (top). For comparison, also shown is the gait produced by IS-MPC when the disturbance is known (bottom).}
\label{fig:lipSim3}
\end{figure}

The disturbance signal used in the third simulation is $d(t) =0.2+0.15\sin(2\pi t)$~m/s$^2$, which includes a sinusoidal term that varies more rapidly. Pure observer-based IS-MPC fails in this case because it becomes unfeasible (results not shown). However, as shown in Fig.~\ref{fig:lipSim3}, feasibility is recovered if the ZMP constraints are restricted progressively along the control horizon. This stratagem, inspired by~\cite{ChRoZa:01}, is effective in this case because the ZMP constraint restriction can be shown to be beneficial for feasibility. Note how the gait is quite different from that produced by IS-MPC if the disturbance is known, also shown in Fig.~\ref{fig:lipSim3}; this is consistent with the reduced admissible regions for the ZMP.

\begin{figure}[t]
\centering
\includegraphics[height=0.7\columnwidth]{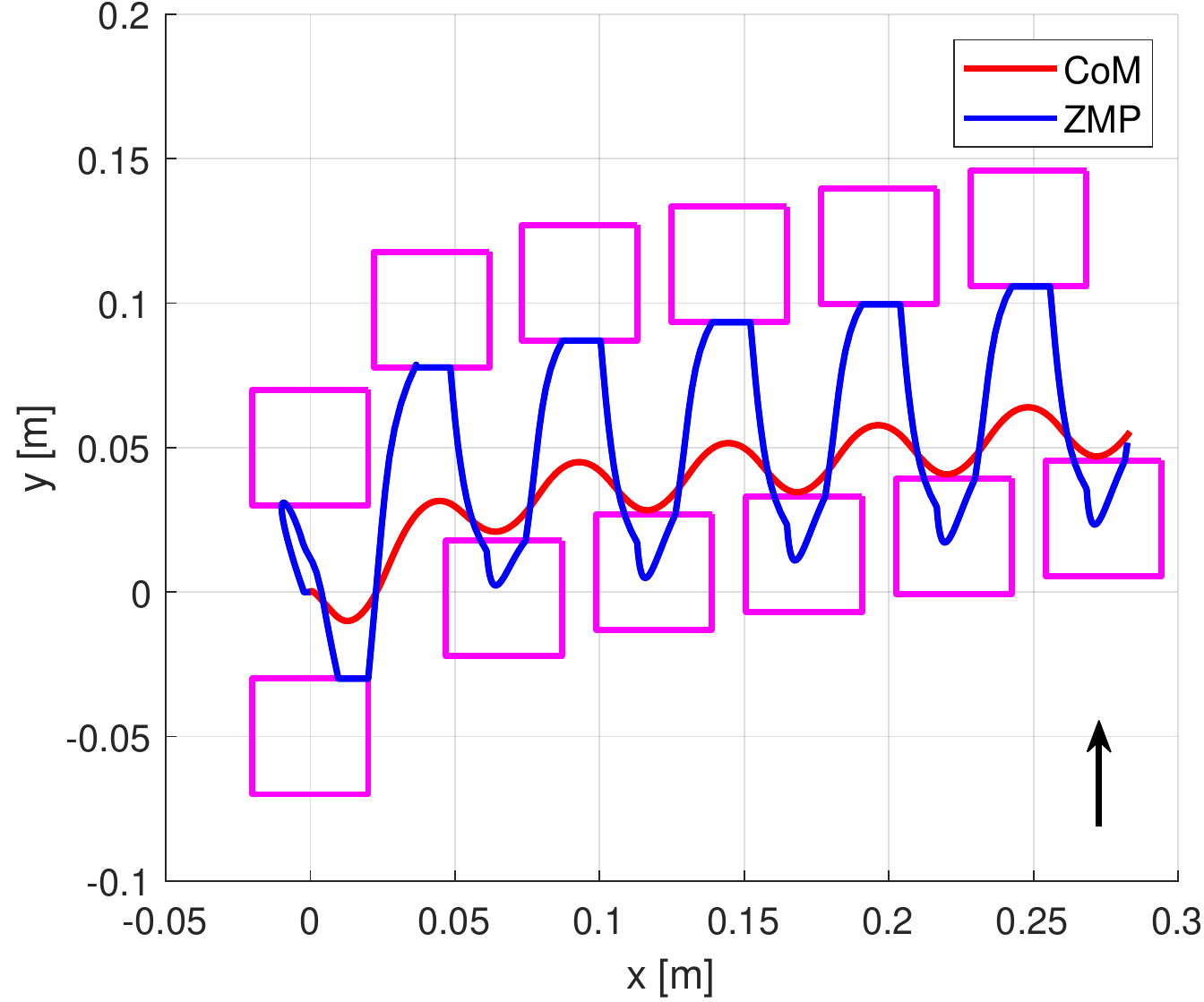}
\caption{Gait generation in the presence of an unknown constant disturbance on the CoM: result of observer-based IS-MPC with automatic footstep placement. Note the direction of the disturbance.}
\label{fig:lipAfp}
\end{figure}

Finally, we have simulated an observer-based IS-MPC scheme with automatic footstep placement in the presence of a constant disturbance $\bar d = 0.4$~m/s$^2$ acting along the $y$ axis. 
To perform AFS, the footstep positions are added to the decision variables of the MPC, while the cost function is modified by including a term for tracking a reference velocity of the CoM~\cite{ScCoDeLaOr:16}, in this case 0.1~m/s along the $x$ axis. In the resulting gait, shown in Fig.~\ref{fig:lipAfp}, one observes the expected displacement of the footsteps due to the disturbance.


\subsection{Dynamic simulations on NAO}

As a further validation step, we have performed dynamic simulations of the proposed method for a NAO humanoid robot in DART (Dynamic Animation and Robotics Toolkit). The qpOASES library was used to solve the QP. The robot and gait parameters are the same as in the previous simulations, except for $\delta = 0.05$~s and $C=20$. 

\begin{figure}[t]
    \centering
    \includegraphics[width=4.2 cm]{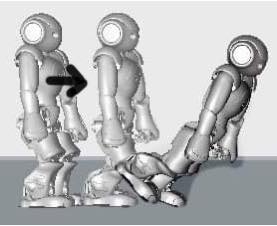}
    \includegraphics[width=3.89 cm]{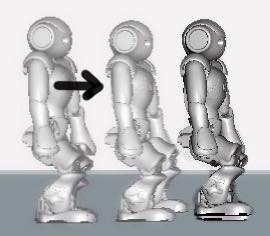}
    \vskip 0.4 cm
    \includegraphics[width=\columnwidth]{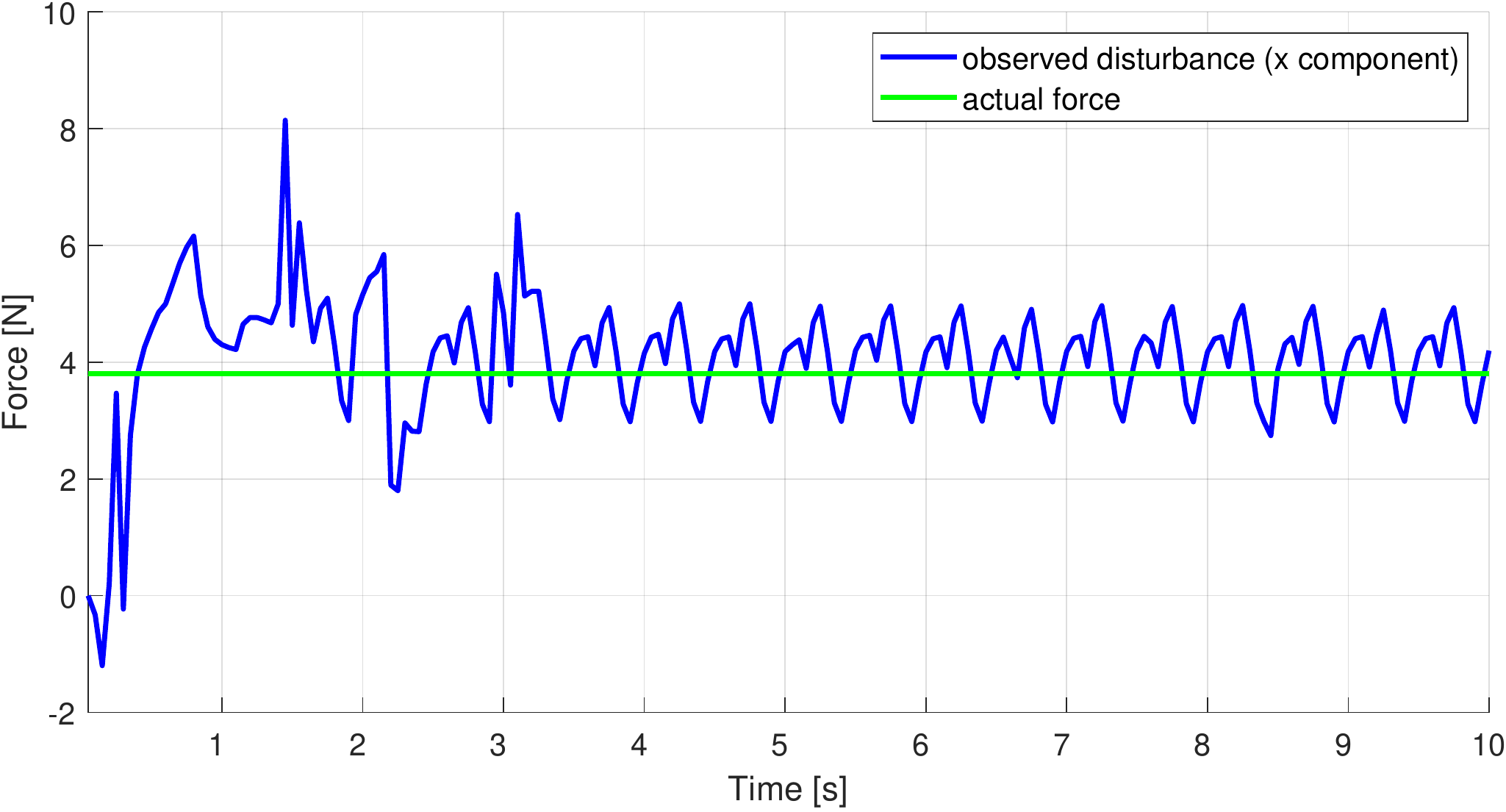}
    \caption{NAO dynamic simulation in the presence of an unknown constant force acting on the CoM. With IS-MPC, the robot is unable to maintain balance (top left). With observer-based IS-MPC, the robot successfully counteracts the disturbance (top right). Also shown is the observed force against the actual force (bottom).}
    \label{Falling}
 \end{figure}    
    
\begin{figure} [t]    
    \includegraphics[width=\columnwidth]{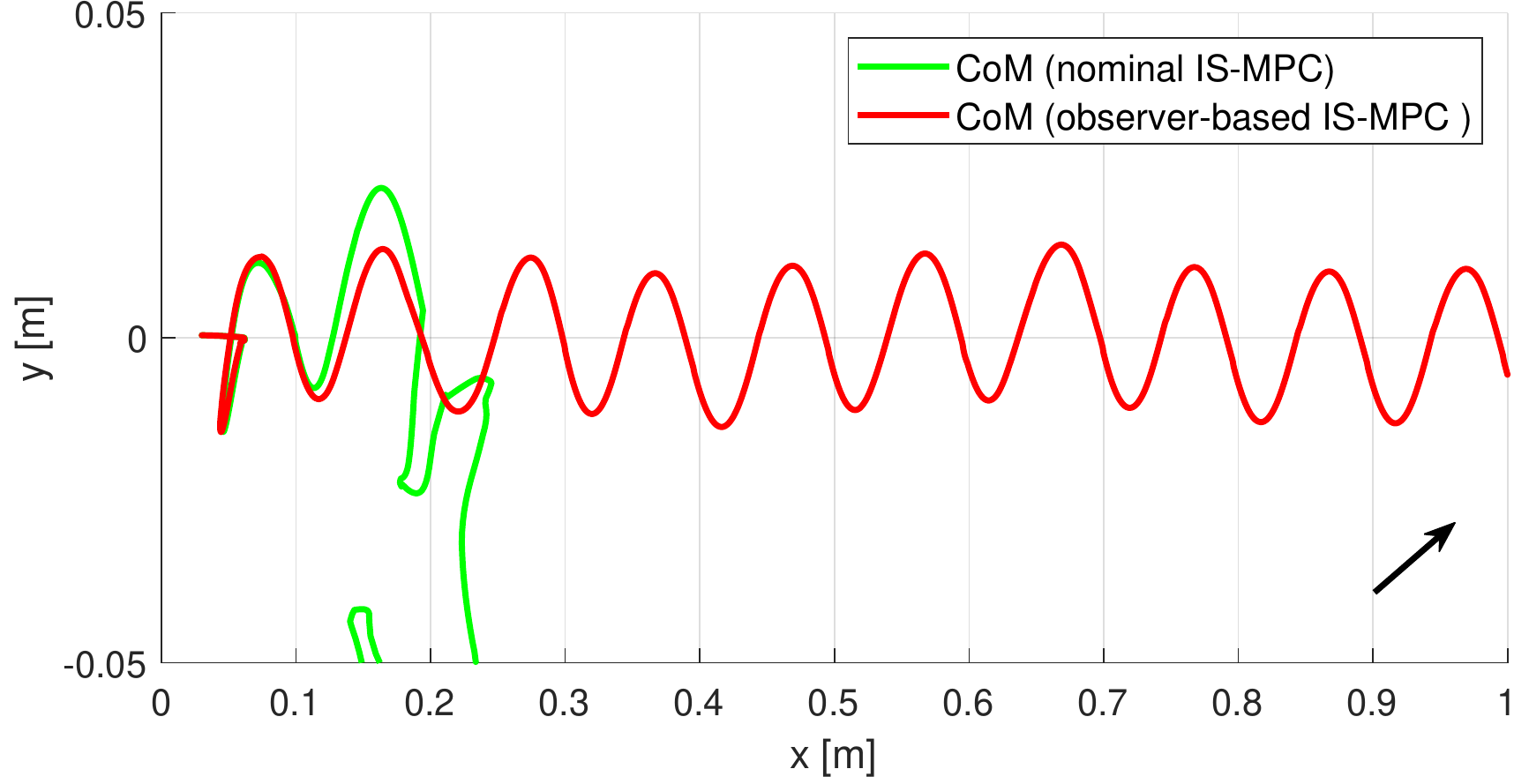}
    \caption{NAO dynamic simulation in the presence of a slowly-varying force acting on the CoM. With nominal IS-MPC, the robot is unable to maintain balance, whereas with observer-based IS-MPC a stable gait is achieved.}
    \label{fig:Against}
\end{figure} 

In the first dynamic simulation, a constant external force of $3.8$~N along the sagittal axis is applied to the robot CoM. As shown in Fig.~\ref{Falling}, the robot falls when nominal IS-MPC is used, whereas observer-based IS-MPC allows to counteract the disturbance successfully, producing the aforementioned effect of leaning against the force. An interesting aspect of this simulation, clearly shown in the bottom plot, is that the observer does not estimate only the constant force, because it also reacts to dynamic effects that are not modeled in the LIP (see~(\ref{eq:mism}).

In the second simulation, $F_{\rm ext}=2+3.8\sin(0.45\pi t)$~N includes a slow sinusoidal component. Figure~\ref{fig:Against} shows a comparison between the CoM trajectories generated by nominal vs.\ observer-based IS-MPC. Once again, the first fails while the second is able to maintain balance while walking.

\begin{figure}[t]
    \centering
    \includegraphics[width=0.68\columnwidth]{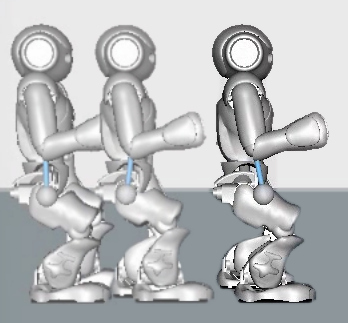}
    \vskip 0.3 cm
    \includegraphics[width=\columnwidth]{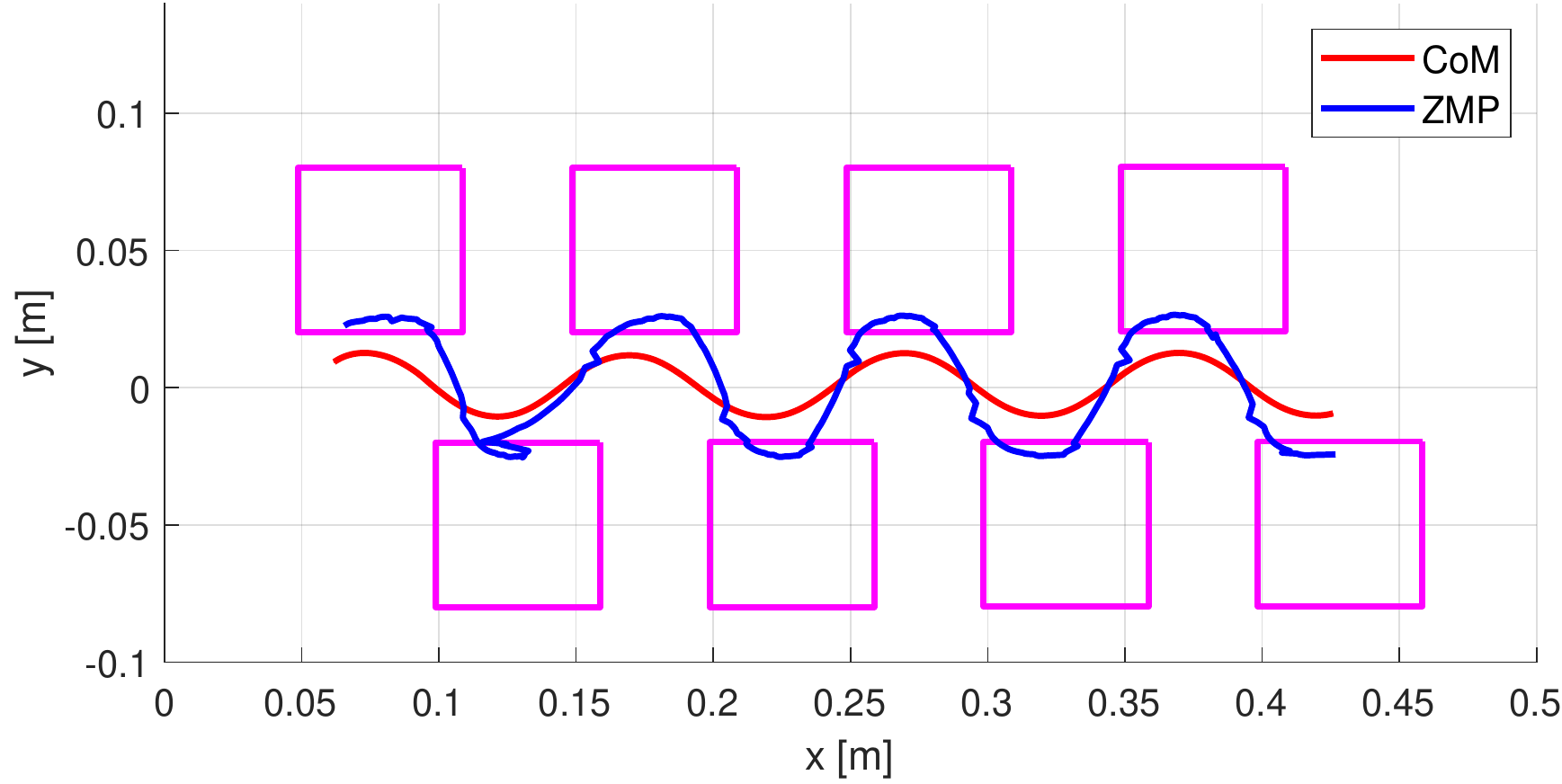}
    \vskip 0.3 cm
    \includegraphics[width=\columnwidth]{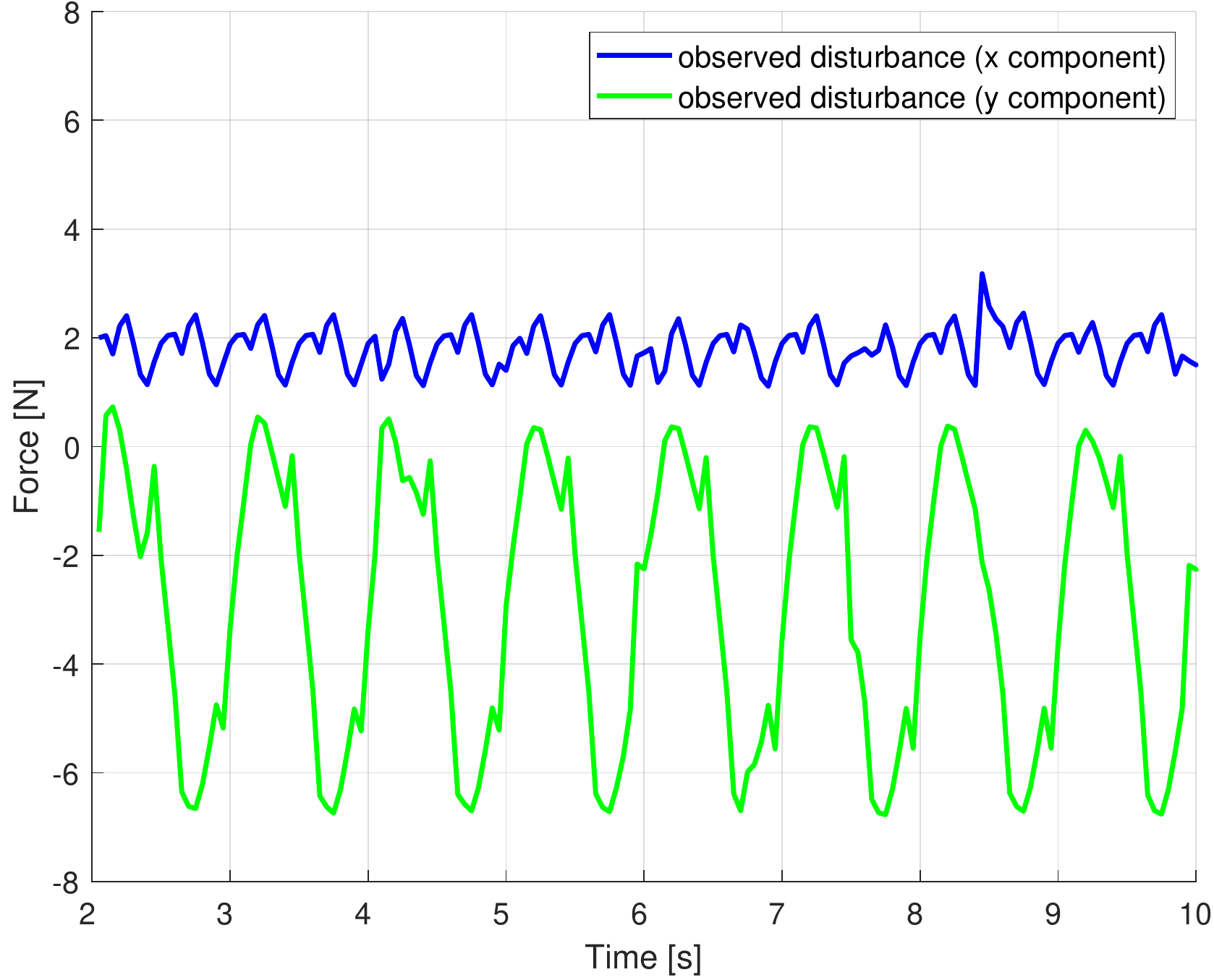}
    \caption{NAO dynamic simulation in the presence of an oscillating mass attached to the arm (top). Using observer-based IS-MPC, a stable gait is achieved (center). Also shown are the observed disturbances along the two axes (bottom).}
    \label{fig:Pendulum}
\end{figure}

In a third simulation, shown in Fig.~\ref{fig:Pendulum}, we considered a more realistic scenario where the disturbance is not directly applied to the robot CoM. In particular, the humanoid is carrying attached to its arm a $0.2$~kg pendulum. This could represent, e.g., an oscillating weight such as a shopping bag. Thanks to the use of observer-based IS-MPC, the robot successfully counteracts the disturbance.

Movie clips of the above dynamic simulations are shown in the video attachment.


\section{Conclusions}
\label{sec:conclusions}

We have presented an extension of our previously proposed IS-MPC scheme which is able to generate stable humanoid gaits in the presence of persistent disturbances. To this end, it incorporates a disturbance observer providing an estimate which is then used to correct appropriately the stability constraints.  The resulting observer-based IS-MPC scheme was validated via simulations on a LIP model and a NAO humanoid, showing successful gait generation for a wide range of applied disturbances.

Future work will include:

\begin{itemize}

\item experimental validation of the proposed scheme (note that the computational load of observer-based IS-MPC is virtually the same of the standard IS-MPC, so that a real-time implementation is possible);

\item adaptation to more general classes of disturbances;

\item a study of the conditions  for recursive feasibility of the observer-based IS-MPC algorithm.

\end{itemize}


\bibliographystyle{IEEEtran}
\bibliography{IEEEabrv,robustStableMPC}

\end{document}